\documentclass[runningheads,11pt]{llncs}
\usepackage{pgf}
\usepackage {epic}
\usepackage {eepic}
\usepackage{latexsym}
\usepackage{mathrsfs}
\usepackage{amssymb}
\usepackage{graphics}

\authorrunning{N.~Alon, F.~V.Fomin, ~G.~Gutin, M.~Krivelevich, and S.~Saurabh}

\usepackage{amsmath}
\usepackage{eepic}
\usepackage {epic}
\usepackage {eepic}
\usepackage{graphics}
\usepackage{amssymb}
\usepackage{url}
\usepackage{amssymb}
\usepackage{mathrsfs}
\newtheorem{defn}[theorem]{Definition}
\newtheorem{fact}[theorem]{Fact}
\newtheorem{rem}[theorem]{Remark}
\usepackage{xspace}

\usepackage{vmargin}

\setmarginsrb{3.8cm}{3.9cm}{3.9cm}{3.9cm}{0.5cm}{0.5cm}{0.5cm}{0.5cm}

\title{Better Algorithms and Bounds for Directed Maximum Leaf Problems}

\author{
Noga Alon\inst{1} \and Fedor V. Fomin\inst{2}  \and Gregory
Gutin\inst{3} \and Michael Krivelevich\inst{1} \and Saket
Saurabh\inst{2,4}}

\institute{Department of Mathematics, Tel Aviv University\\ Tel Aviv
69978, Israel\\\email{\{nogaa,krivelev\}@post.tau.ac.il}\and
Department of Informatics, University of Bergen\\ POB 7803, 5020
Bergen,
Norway\\
 \email{\{fedor.fomin,saket\}@ii.uib.no}
\and Department of Computer Science\\
 Royal Holloway, University of London\\
Egham, Surrey TW20 0EX, UK\\
\email{gutin@cs.rhul.ac.uk} \and The Institute of Mathematical
Sciences\\ Chennai, 600 017, India\\ \email{saket@imsc.res.in}}

\begin{document}
\date{}
\maketitle

\begin{abstract}
The {\sc Directed Maximum Leaf Out-Branching}  problem is  to find
an out-branching (i.e. a rooted oriented spanning tree) in a given
digraph with the maximum number of leaves. In this paper, we improve
known parameterized algorithms and combinatorial bounds on the
number of leaves in out-branchings. We show that
\smallskip
\begin{itemize}
\item  every strongly connected digraph $D$ of
order $n$ with minimum in-degree at least 3 has an out-branching
with at least $(n/4)^{1/3}-1$ leaves;
\item if a strongly connected digraph $D$
does not contain an out-branching with $k$ leaves, then the
pathwidth of its underlying graph  is  $O(k\log k)$;
\item  it can be decided in time $2^{O(k\log^2 k)}\cdot n^{O(1)}$ whether
a strongly connected digraph on $n$ vertices has an out-branching
with at least $k$ leaves.
\end{itemize}
\smallskip

All improvements use properties of extremal structures obtained
after applying local search and of some out-branching
decompositions.

\end{abstract}

\section{Introduction}\label{introsec}
Given a digraph $D$, a subdigraph $T$ of $D$ is an {\em out-tree} if
$T$ is an oriented tree with only one vertex $s$ of in-degree zero
(called {\em the root}) and if $T$ is a spanning out-tree, i.e.
$V(T)=V(D)$, then $T$ is called an {\em out-branching} of $D$. The
vertices of $T$ of out-degree zero are called {\em leaves}. The {\sc
Directed Maximum Leaf Out-Branching} problem is to find an
out-branching in a given digraph with the maximum number of leaves.
This problem is a natural generalization of the intensively studied
{\sc Maximum Leaf Spanning Tree} problem on connected undirected
graphs
\cite{bonsmaLNCS2747,DingJS01,estivill,fominGK06,FellowsMRS00,GalbiatMM97,GriggsW92,LuR98,Solis-Oba98}.
Unlike its undirected counterpart which has attracted a lot of
attention in all algorithmic paradigms like approximation
algorithms~\cite{GalbiatMM97,LuR98,Solis-Oba98}, parameterized
algorithms~\cite{bonsmaLNCS2747,estivill,FellowsMRS00}, exact
exponential time algorithms~\cite{fominGK06} and also combinatorial
studies~\cite{DingJS01,GriggsW92,KleitmanW91,LinialS87}, the {\sc
Directed Maximum Leaf Out-Branching} problem has largely been
neglected until recently.

In \cite{alonLNCS4596} we initiated algorithmic and combinatorial
study of {\sc Directed Maximum Leaf Out-Branching} and obtained, as
the main result of the paper, the first fixed parameter tractable
algorithms for the problem on strongly connected digraphs and
acyclic digraphs based on various combinatorial lemmas. In this
paper we continue our investigation of the {\sc Directed Maximum
Leaf Out-Branching} (DMLOB) and obtain several improved
parameterized algorithms for the problem as well as combinatorial
results regarding the number of leaves possible in an out-branching
of a digraph based on completely new approaches and ideas which are
interesting in its own and could be useful for solving other
problems on digraphs.

In parameterized algorithms, for decision problems with input size
$n$, and a parameter $k$, the goal of is to design an algorithm with
runtime $f(k)n^{O(1)}$, where $f$ is a function of $k$ alone. (For
DMLOB such a parameter is the number of leaves in the out-tree.)
Problems having such an algorithm are said to be fixed parameter
tractable (FPT). The book by Downey and Fellows \cite{downey1999}
provides an introduction to the topic of parameterized complexity.
For recent developments see the books by Flum and Grohe
\cite{FlumGrohebook} and by Niedermeier \cite{Niedermeierbook06}.

The parameterized version of DMLOB is defined as follows: Given a
digraph $D$ and a positive integral parameter $k$, does there exist
an out-branching with at least $k$ leaves? We denote the
parameterized versions of DMLOB by $k$-DMLOB. If in the above
definition we do not insist for an out-branching and ask whether
there exists an out-tree with at least $k$ leaves, we get
parameterized {\sc Directed Maximum Leaf Out-Tree} problem (denoted
$k$-DMLOT).


\smallskip

In this paper we obtain the following new algorithmic and
combinatorial results on $k$-DMLOB for strongly connected digraphs
and acyclic digraphs. Before we go any further we remark that the
algorithmic results presented here also hold for {\em all digraphs}
if we consider $k$-DMLOT rather than $k$-DMLOB. However, we mainly
restrict ourselves to $k$-DMLOB for clarity and the harder
challenges it poses, and we briefly consider $k$-DMLOT only in the
last section.

\medskip
\noindent\textbf{Faster Algorithm.} We design a new algorithm which
decides in time $2^{O(k\log^2 k)}\cdot n^{O(1)}$ whether a strongly
connected digraph on $n$ vertices has an out-branching with at least
$k$ leaves (Corollary~\ref{cor:mainresult}). On acyclic graphs we
can solve the problem even faster, in time $2^{O(k\log k)}\cdot
n^{O(1)}$ (Corollary~\ref{cor:acyclic}). These are significant
improvements over running time $2^{O(k^2\log k)}\cdot n^{O(1)}$ for
both classes of digraphs obtained in \cite{alonLNCS4596}. The
improvements do not result from a careful tuning of the algorithm
from \cite{alonLNCS4596} but from several novel ideas. In
particular, we use local search and specific tree partition
arguments. While local search is a widely used technique in
heuristics and approximation algorithms (see, e.g.,
\cite{AartsL97-Lo}) we are not aware of its applications in
parameterized complexity. We find it to be of independent interest.

\noindent \textbf{Combinatorial bounds.} Kleitman and West
\cite{KleitmanW91} and Linial and Sturtevant \cite{LinialS87} showed
that every connected undirected graph $G$ on $n$ vertices with
minimum degree at least $3$ has a spanning tree with at least $n/4 +
2$ leaves. In \cite{alonLNCS4596} we proved an analogue of this
result for directed graphs: every strongly connected digraph $D$ of
order $n$ with minimum in-degree at least 3 has an out-branching
with at least $(n/2)^{1/5}-1$ leaves. In this paper
(Theorem~\ref{main1}), we improve this bound to  $(n/4)^{1/3}-1$. We
do not know whether the last bound is tight, however we show that
there are strongly connected digraphs with minimum in-degree 3 in
which every out-branching has at most $O(\sqrt{n})$ leaves (Theorem
\ref{exampleth}).
Another parallel between the worlds of directed and undirected
graphs established in this paper (and used intensively in the
algorithmic part) is the relation between the number of leaves in a
maximum leaf out-branching in a digraph $D$ and the pathwidth of its
underlying graph. It is easy to check (see, e.g.,
\cite{BienstockRST91}), that every connected undirected graph of
pathwidth at least $k$, contains a spanning tree with at least $k$
leaves. We show (Theorem~\ref{mainth}) that if a strongly connected
digraph $D$ does not contain an out-branching with $k$ leaves, then
the pathwidth of its underlying graph  is  $O(k\log k)$.


\section{Preliminaries}
\label{prelim} Let $D$ be a digraph. By $V(D)$ and $A(D)$ we
represent the vertex set and arc set of $D$, respectively. An {\em
oriented graph} is a digraph with no directed 2-cycle. Given a
subset $V'\subseteq V(D)$ of a digraph $D$, let $D[V']$ denote the
digraph induced on $V'$. The {\em underlying undirected graph}
$UN(D)$ of $D$ is obtained from $D$ by omitting all orientations of
arcs and by deleting one edge from each resulting pair of parallel
edges. The {\em connectivity components} of $D$ are the subdigraphs
of $D$ induced by the vertices of components of $UN(D)$. A digraph
$D$ is {\em strongly connected} if, for every pair $x,y$ of vertices
there are directed paths from $x$ to $y$ and from $y$ to $x.$ A
maximal strongly connected subdigraph of $D$ is called a {\em strong
component}. A vertex $u$ of $D$ is an {\em in-neighbor} ({\em
out-neighbor}) of a vertex $v$ if $uv\in A(D)$ ($vu\in A(D)$,
respectively). The {\em in-degree} $d^-(v)$ ({\em out-degree}
$d^+(v)$) of a vertex $v$ is the number of its in-neighbors
(out-neighbors).

We denote by $\ell(D)$ the maximum number of leaves in an out-tree
of a digraph $D$ and by $\ell_s(D)$ we denote the maximum possible
number of leaves in an out-branching of a digraph $D$. When $D$ has
no out-branching, we write $\ell_s(D)=0$. The following simple
result gives necessary and sufficient conditions for a digraph to
have an out-branching. This assertion allows us to check whether
$\ell_s(D)>0$ in time $O(|V(D)|+|A(D)|)$.

\begin{proposition}[\cite{bang2000}]\label{iffoutb}
A digraph $D$ has an out-branching if and only if $D$ has a unique
strong component with no incoming arcs.
\end{proposition}

Let $P=u_1u_2\ldots u_q$ be a directed path in a digraph $D$. An arc
$u_iu_j$ of $D$ is a {\em forward} ({\em backward}) {\em arc for}
$P$ if $i\le j-2$ ($j<i$, respectively). Every backward arc of the
type $v_{i+1}v_i$ is called {\em double}.

For a natural number $n$, $[n]$ denotes the set $\{1,2,\ldots ,n\}.$

A {\em tree decomposition} of an (undirected) graph $G$ is a pair
$(X,U)$ where $U$ is a tree whose vertices we will call {\em nodes}
and $X=(\{X_{i} \mid i\in V(U)\})$ is a collection of subsets of
$V(G)$ such that
\begin{enumerate}
\item $\bigcup_{i \in V(U)} X_{i} = V(G)$,

\item for each edge $\{v,w\} \in E(G)$, there is an $i\in V(U)$
such that $v,w\in X_{i}$, and

\item for each $v\in V(G)$ the set of nodes $\{ i \mid v \in X_{i}
\}$ forms a subtree of $U$.
\end{enumerate}
The {\em width} of a tree decomposition $(\{ X_{i} \mid i \in V(U)
\}, U)$ equals $\max_{i \in V(U)} \{|X_{i}| - 1\}$. The {\em
treewidth} of a graph $G$ is the minimum width over all tree
decompositions of $G$.

If in the definitions of a tree decomposition and treewidth we
restrict $U$ to be a tree with all vertices of degree at most $2$
(i.e., a path) then we have the definitions of path decomposition
and pathwidth. We use the notation $tw(G)$ and $pw(G)$ to denote the
treewidth and the pathwidth of a graph $G$.

We also need an equivalent definition of pathwidth in terms of
vertex separators with respect to a linear ordering of the vertices.
Let $G$ be a graph and let $\sigma=(v_1,v_2,\ldots ,v_n)$ be an
ordering of $V(G)$. For $j\in [n]$ put $V_j =\{v_i:\ i\in [j]\}$ and
denote by $\partial V_j$ all vertices of $V_j$ that have neighbors
in $V\setminus V_j.$ Setting $ vs(G,\sigma) = \max_{i\in [n]}
|\partial V_i | , $ we define the \emph{vertex separation}  of  $G$
as
\[
vs(G) = \min \{vs(G,\sigma) \colon \sigma \mbox{ is an ordering of }
V(G)\}.
\]

The following assertion is well-known.  It follows directly from the
results of Kirousis and Papadimitriou \cite{KirousisP85} on interval
width  of a graph, see also \cite{Kinnersley92}.

\begin{proposition}[\cite{Kinnersley92,KirousisP85}]\label{sovp_pw_vs}
For any graph $G$, $vs(G)=pw(G)$.
\end{proposition}



\section{Locally Optimal Out-Trees}
Our improved parameterized algorithms are based on finding locally
optimal out-branchings. Given a digraph, $D$ and an out-branching
$T$, we call a vertex {\em leaf}, {\em link} and {\em branch} if its
out-degree in $T$ is $0$, $1$ and $\geq 2$ respectively. Let
$S^{+}_{\geq 2 }(T)$ be the set of branch vertices, $S^{+}_{1 }(T)$
be the set of link vertices and $L(T)$ be the set of leaves in the
tree $T$. Let $\mathscr{P}_2(T)$ be the set of maximal paths
consisting of link vertices. By $p(v)$ we denote the {\em parent} of
a vertex $v$ in $T$; $p(v)$ is the unique in-neighbor of $v.$ We
call a pair of vertices $u$ and $v$ {\em siblings} if they do not
belong to the same path from the root $r$ in $T$. We start with the
following well known and easy to observe facts.
\begin{fact}
 $|S^{+}_{\geq 2 }(T)| \leq |L(T)|-1$.
\end{fact}

\begin{fact}
 $|\mathscr{P}_2(T)| \leq 2 |L(T)|-1$.
\end{fact}

Now we define the notion of local exchange which is intensively used
in our proofs.
\begin{defn}
{\sc $\ell$-Arc Exchange ($\ell$-AE) optimal out-branching:} An
out-branching $T$ of  a directed graph $D$ with $k$ leaves is
$\ell$-AE optimal if
 for all arc subsets $F\subseteq A(T)$ and $X \subseteq A(D)-A(T)$ of size $\ell$,
 $(A(T)\setminus F) \cup X$  is either not an out-branching,  or an out-branching
with $\leq k$ leaves.
 In other words, $T$ is $\ell$-AE optimal if it can't be turned into an
out-branching with more leaves by exchanging
 $\ell$ arcs.
\end{defn}
Let us remark, that for every fixed $\ell$, an $\ell$-AE optimal
out-branching can be obtained in polynomial time. In our proofs we
use only $1$-AE optimal out-branchings. We need the following simple
properties of $1$-AE optimal out-branchings.
\begin{lemma}
\label{char1ae} Let $T$ be an $1$-AE optimal  out-branching rooted
at $r$ in a digraph $D$. Then the following holds:
\begin{itemize}
\item[(a)] For every pair of siblings $u,v\in V(T)\setminus L$ with  $d^{+}_T(p(v))
= 1$, there is no arc $e=(u,v)\in A(D)\setminus A(T)$;
\item[(b)] For every pair of vertices $u,v \notin L$, $d^{+}_T(p(v)) = 1$, which are
 on the same path from the root with $dist(r,u) < dist(r,v)$
there is no arc $e=(u,v)\in A(D)\setminus A(T)$ (here $dist(r,u)$ is
the distance to $u$ in $T$ from the root $r$);
\item[(c)] There is no arc $(v,r)$, $v \notin L$ such that the
directed cycle formed by the $(r,v)$-path and the arc $(v,r)$
contains a  vertex $x$ such that $d^{+}_T(p(x)) = 1$.
\end{itemize}
\end{lemma}

\section{Combinatorial Bounds}

We start with a lemma that allows us to obtain lower bounds on
$\ell_s(D)$.
\begin{lemma}\label{comblemma}
Let $D$ be a oriented graph of order $n$ in which every vertex is of
in-degree 2 and let $D$ have an out-branching. If $D$ has no
out-tree with $k$ leaves, then $n\le 4k^3.$
\end{lemma}
\begin{proof} Let us assume that $D$ has no out-tree with $k$ leaves. Consider
an out-branching $T$ of $D$ with $p<k$ leaves  which is $1$-AE
optimal. Let $r$ be the root of $T$.

We will bound the number $n$ of vertices in $T$ as follows. Every
vertex of $T$ is either a leaf, or a branch vertex, or a link
vertex. By Facts~1 and 2 we already have bounds on the number of
leaf and branch vertices as well as the number of maximal paths
consisting of link vertices. So to get an upper bound on $n$ in
terms of $k$, it suffices to bound the length of each maximal path
consisting of link vertices. Let us consider such a  path $P$ and
let $x,y$ be the first and last vertices of $P$, respectively.

The vertices of $V(T)\setminus V(P)$ can be partitioned into four
classes as follows:
\begin{itemize}\item[$(a)$] {\sf ancestor vertices}: the vertices which appear
before $x$ on the $(r,x)$-path of $T$;
\item[$(b)$]
{\sf descendant vertices }: the vertices appearing after the
vertices of $P$ on paths of $T$ starting at $r$ and passing through
$y$;
\item[$(c)$]{\sf sink vertices}: the vertices which are leaves but not descendant
vertices;
\item[$(d)$]
{\sf special vertices}: none-of-the-above vertices.
\end{itemize}

Let $P'=P-x$, let $z$ be the out-neighbor of $y$ on $T$ and let
$T_z$ be the subtree of $T$ rooted at $z$. By Lemma~\ref{char1ae},
there are no arcs from special or ancestor vertices to the path
$P'$. Let $uv$ be an arc of $A(D)\setminus A(P')$ such that $v\in
V(P').$ There are two possibilities for $u$: (i) $u\not\in V(P')$,
(ii) $u\in V(P')$ and $uv$ is backward for $P'$ (there are no
forward arcs for $P'$ since $T$ is 1-AE optimal). Note that every
vertex of type (i) is  either a descendant vertex or a sink. Observe
also that the backward arcs for $P'$ form a vertex-disjoint
collection of out-trees with roots at vertices that are not terminal
vertices of backward arcs for $P'$. These roots are terminal
vertices of arcs in which first vertices are descendant vertices or
sinks.

We denote by $\{u_1,u_2,\ldots, u_s\}$ and $\{v_1,v_2,\ldots, v_t\}$
the sets of vertices on $P'$ which have out-neighbors that are
descendant vertices and sinks, respectively. Let the out-tree formed
by backward arcs for $P'$ rooted at $w\in
\{u_1,\ldots,u_s,v_1,\ldots,v_t\}$ be denoted by $T(w)$ and let
$l(w)$ denote the number of leaves in $T(w).$ Observe that the
following is an out-tree rooted at $z$:
$$T_z \cup \{(in(u_1),u_1), \ldots, (in(u_s),u_s)\} \cup \bigcup_{i=1}^{s}T(u_i),$$
where $\{in(u_1),\ldots,in(u_s)\}$ are the in-neighbors of
$\{u_1,\ldots,u_s\}$ on $T_z.$ This out-tree has at least
$\sum_{i=1}^{s}l(u_i)$ leaves and, thus, $\sum_{i=1}^{s}l(u_i)\le
k-1.$ Let us denote the subtree of $T$ rooted at $x$ by $T_x$ and
let $\{in(v_1),\ldots,in(v_t)\}$ be the in-neighbors of
$\{v_1,\ldots,v_t\}$ on $T-V(T_x)$. Then we have following out-tree:
$$(T-V(T_x)) \cup \{(in(v_1),v_1), \ldots, (in(v_t),v_t)\} \cup
\bigcup_{i=1}^{t}T(v_i)$$ with at least $\sum_{i=1}^{t}l(v_i)$
leaves. Thus, $\sum_{i=1}^{t}l(v_i)\le k-1.$

Consider a path $R=v_0v_1\ldots v_r$ formed by backward arcs.
Observe that the arcs $\{v_iv_{i+1}:\ 0\le i\le r-1\}\cup
\{v_jv^+_j:\ 1\le j\le r\}$ form an out-tree with $r$ leaves, where
$v^+_j$ is the out-neighbor of $v_j$ on $P.$ Thus, there is no path
of backward arcs of length more than $k-1$. Every out-tree $T(w)$,
$w\in \{u_1,\ldots,u_s \}$ has $l(w)$ leaves and, thus, its arcs can
be decomposed into $l(w)$ paths, each of length at most $k-1$. Now
we can bound the number of arcs in all the trees $T(w)$, $w\in
\{u_1,\ldots,u_s \}$, as follows: $ \sum_{i=1}^{s}l(u_i)(k-1) \leq
(k-1)^2.$ We can similarly bound the number of arcs in all the trees
$T(w)$, $w\in \{v_1,\ldots,v_s \}$ by $(k-1)^2$. Recall that the
vertices of $P'$ can be either terminal vertices of backward arcs
for $P'$ or vertices in $\{u_1,\ldots,u_s,v_1,\ldots,v_t\}$. Observe
that $s+t\le 2(k-1)$ since $\sum_{i=1}^{s}l(u_i)\le k-1$ and
$\sum_{i=1}^{t}l(v_i)\le k-1.$

Thus, the number of vertices in $P$ is bounded from above by
$1+2(k-1)+2(k-1)^2$. Therefore,
\begin{eqnarray*}
n &=& |L(T)| + |S^{+}_{\geq 2 }(T)|+ |S^{+}_{ 1}(T)|\\
  &=& |L(T)| + |S^{+}_{\geq 2 }(T)|+ \sum_{P \in \mathscr{P}_2(T)} |V(P)|\\
  & \leq & (k-1) + (k-2) + (2k-3) (2k^2-2k+1)\\
  & < & 4k^3.
\end{eqnarray*}
Thus, we conclude that $n\le 4k^3.$ \qed\end{proof}

\begin{theorem}\label{main1}
Let $D$ be a strongly connected digraph with $n$ vertices.
\begin{enumerate}
\item[(a)] If  $D$ is an oriented graph with minimum in-degree at
least 2, then $\ell_s(D)\ge (n/4)^{1/3}-1.$ \item[(b)] If $D$ is a
digraph with minimum in-degree at least 3, then $\ell_s(D)\ge
(n/4)^{1/3}-1.$
\end{enumerate}
\end{theorem}
\begin{proof} Since $D$ is strongly connected, we have $\ell(D)=\ell_s(D)>0.$
Let $T$ be an 1-AE optimal out-branching of $D$ with maximum number
of leaves. (a) Delete some arcs from $A(D)\setminus A(T)$, if
needed, such that the in-degree of each vertex of $D$ becomes 2. Now
the inequality $\ell_s(D)\ge (n/4)^{1/3}-1$ follows from Lemma
\ref{comblemma} and the fact that $\ell(D)=\ell_s(D)$.

(b)  Let $P$ be the path formed in the proof of Lemma
\ref{comblemma}. (Note that $A(P)\subseteq A(T)$.) Delete every
double arc of $P$, in case there are any, and delete some more arcs
from $A(D)\setminus A(T)$, if needed, to ensure that the in-degree
of each vertex of $D$ becomes 2. It is not difficult to see that the
proof of Lemma \ref{comblemma} remains valid for the new digraph
$D$. Now the inequality $\ell_s(D)\ge (n/4)^{1/3}-1$ follows from
Lemma \ref{comblemma} and the fact that $\ell(D)=\ell_s(D)$. \qed
 \end{proof}

\begin{rem}\label{mainremark}
It is easy to see that Theorem \ref{main1} holds also for acyclic
digraphs $D$ with $\ell_s(D)>0$.
\end{rem}

While we do not know whether the bounds of Theorem \ref{main1} are
tight, we can show that no linear bounds are possible. The following
result is formulated for Part (b) of Theorem \ref{main1}, but a
similar result holds for Part (a) as well.

\begin{theorem}\label{exampleth}
For each $t\ge 6$ there is a strong digraph $H_t$ of order $n=t^2+1$
with minimum in-degree 3 such that $0<\ell_s(H_t)=O(t).$
\end{theorem}
\begin{proof} Let $V(H_t)=\{r\}\cup \{u^i_1,u^i_2,\ldots ,u^i_{t}~|\ i\in
[t]\}$ and
\begin{eqnarray*}
A(H_t)& = & \left\{u^i_ju^i_{j+1},u^i_{j+1}u^i_j~|~\ i\in [t], j\in
\{0,1,\ldots ,t-3\}\right\} \\
& & \bigcup \left\{u^i_ju^i_{j-2}~|~ i\in [t], j\in
\{3,4,\ldots ,t-2\} \right \} \\
& & \bigcup \left\{u^i_ju^i_q ~|~\ i\in [t], t-3\le j\neq q\le
t\right\},
\end{eqnarray*}
where $u^i_0=r$ for every $i\in [t].$ It is easy to check that
$0<\ell_s(H_t)=O(t).$\qed
\end{proof}

%

\section{Decomposition Algorithms}

\begin{theorem}\label{dagth} Let $D$ be an acyclic
digraph with a single vertex of in-degree zero. Then either
$\ell_s(D)\ge k$ or the underlying undirected graph of $D$ is of
pathwidth at most $4k$ and we can obtain this path decomposition in
polynomial time.
\end{theorem}
\begin{proof}
Assume that $\ell_s(D)\le k-1$. Consider a $1$-AE optimal
out-branching $T$ of $D$. Notice that $|L(T)|\le k-1.$ Now remove
all the leaves and branch vertices from the tree $T$. The remaining
vertices form maximal directed paths consisting of link vertices.
Delete the first vertices of all paths. As a result we obtain a
collection $\cal Q$ of directed paths. Let $H=\cup_{P\in {\cal
Q}}P$. We will show that every arc $uv$ with $u,v\in V(H)$ is in
$H.$

Let $P'\in \cal Q$. As in the proof of Lemma \ref{comblemma}, we see
that there are no forward arcs for $P'$. Since $D$ is acyclic, there
are no backward arcs for $P'.$ Suppose $uv$ is an arc of $D$ such
that $u\in R'$ and $v\in P'$, where $R'$ and $P'$ are distinct paths
from $\cal Q$. As in the proof of Lemma \ref{comblemma}, we see that
$u$ is either a sink or a descendent vertex for $P'$ in $T$. Since
$R'$ contains no sinks of $T$, $u$ is a descendent vertex, which is
impossible as $D$ is acyclic. Thus, we have proved that
$pw(UN(H))=1.$

Consider a path decomposition of $H$ of width 1. We can obtain a
path decomposition of $UN(D)$ by adding all the vertices of
$L(T)\cup S^{+}_{\geq 2 }(T)\cup F(T)$, where $F(T)$ is the set of
first vertices of maximal directed paths consisting of link vertices
of $T$, to each of the bags of a path decomposition of $H$ of width
1. Observe that the pathwidth of this decomposition is  bounded from
above by
$$|L(T)| + |S^{+}_{\geq 2 }(T)| + |F(T)|+1
\leq (k-1) + (k-2) + (2k-4)+1 \leq 4k -6.$$ The bounds on the
various sets in the inequality above follows from Facts $1$ and $2$.
This proves the theorem.
 \qed\end{proof}

\begin{corollary}\label{cor:acyclic}
For acyclic digraphs, the problem $k$-DMLOB can solved in time
$2^{O(k\log k)}\cdot n^{O(1)}$.
\end{corollary}
\begin{proof}The proof of Theorem~\ref{dagth} can be easily turned into a
polynomial time algorithm to either build an out-branching of $D$
with at least $k$ leaves or to show that $pw(UN(D))\le 4k$ and
provide the corresponding path decomposition. A simple dynamic
programming over the path decomposition gives us an algorithm of
running time $2^{O(k\log k)}\cdot n^{O(1)}$.\qed
 \end{proof}

The following lemma is well known, see, e.g., \cite{chung1990}.

\begin{lemma}
\label{treesep} Let $T=(V,E)$ be an undirected tree and let $w ~:~V
\rightarrow \mathbb{R}^{+} \cup \{0\}$ be a weight function on its
vertices. There exists a vertex $v\in T$ such that the weight of
every subtree $T'$ of $T-v$ is at most $w(T)/2$, where $w(T)=\sum_{v
\in V} w(v)$.
\end{lemma}

Let $D$ be a digraph with $\ell_s(D)=\lambda$ and let $T$ be an
out-branching of $D$ with $\lambda$ leaves. Consider the following
decomposition of $T$ (called a $\beta$-{\em decomposition}) which is
useful in the proof of Theorem~\ref{mainth}.

Assign weight 1 to all leaves of $T$ and weight 0 to all non-leaves
of $T$. By Lemma \ref{treesep}, $T$ has a vertex $v$ such that each
component of $T-v$ has at most $\lambda/2+1$ leaves (if $v$ is not
the root and its in-neighbor $v^-$ in $T$ is a link vertex, then
$v^-$ becomes a new leaf). Let $T_1,T_2,\ldots , T_s$ be the
components of $T-v$ and let $l_1,l_2,\ldots ,l_s$ be the numbers of
leaves in the components. Notice that $\lambda\le \sum_{i=1}^sl_i\le
\lambda+1$ (we may get a new leaf). We may assume that $l_s\le
l_{s-1}\le \cdots \le l_1\le \lambda/2+1.$ Let $j$ be the first
index such that $\sum_{i=1}^{j}l_i \geq \frac{\lambda}{2}+1.$
Consider two cases: (a) $l_{j}\le (\lambda+2)/4$ and (b) $l_{j}>
(\lambda+2)/4$. In Case (a), we have
$$ \frac{\lambda+2}{2}\leq \sum_{i=1}^{j}l_i \leq \frac{3(\lambda+2)}{4} \mbox{ and }\frac{\lambda-6}{4}
\le \sum_{i=j+1}^{s}l_i \le  \frac{\lambda}{2}.$$ In Case (b), we
have $j=2$ and
$$ \frac{\lambda+2}{4}\le l_1\le  \frac{\lambda+2}{2} \mbox{ and }\frac{\lambda-2}{2}
\leq \sum_{i=2}^{s}l_i \le \frac{3\lambda+2}{4}.$$

Let $p=j$ in Case (a) and $p=1$ in Case (b). Add to $D$ and $T$ a
{\em copy} $v'$ of $v$ (with the same in- and out-neighbors). Then
the number of leaves in each of the out-trees
$$T'=T[\{v\}\cup (\cup_{i=1}^pV(T_i))] \mbox{ and } T''=T[\{v'\}\cup
(\cup_{i=p+1}^sV(T_i))]$$ is between $\lambda(1+o(1))/4$ and
$3\lambda(1+o(1))/4$. Observe that the vertices of $T'$ have at most
$\lambda+1$ out-neighbors in $T''$ and the vertices of $T''$ have at
most $\lambda+1$ out-neighbors in $T'$ (we add 1 to $\lambda$ due to
the fact that $v$ `belongs' to both $T'$ and $T''$).

Similarly to deriving $T'$ and $T''$ from $T$, we can obtain two
out-trees from $T'$ and two out-trees from $T''$ in which the
numbers of leaves are approximately between a quarter and three
quarters of the number of leaves in $T'$ and $T''$, respectively.
Observe that after $O(\log \lambda)$ `dividing' steps, we will end
up with $O(\lambda)$ out-trees with just one leaf, i.e., directed
paths. These paths contain $O(\lambda)$ copies of vertices of $D$
(such as $v'$ above). After deleting the copies, we obtain a
collection of $O(\lambda)$ disjoint directed paths covering $V(D)$.

\begin{theorem}\label{mainth} Let $D$ be a strongly connected digraph.
Then either $\ell_s(D)\ge k$ or the underlying undirected graph of
$D$ is of pathwidth $O(k \log k)$.
\end{theorem}
\begin{proof}
We may assume that $\ell_s(D)<k$. Let $T$ be be a 1-AE optimal
out-branching. Consider a $\beta$-decomposition of $T$. The
decomposition process can be viewed as a tree $\cal T$ rooted in a
node (associated with) $T$. The sons of $T$ in $\cal T$ are nodes
(associated with) $T'$ and $T''$; the leaves of $\cal T$ are the
directed paths of the decomposition. The {\em first layer} of $\cal
T$ is the node $T$, the {\em second layer} are $T'$ and $T''$, the
{\em third layer} are sons of $T'$ and $T''$, etc. In what follows,
we do not distinguish between a node $Q$ of $\cal T$ and the tree
associated with the node. Assume that $\cal T$ has $t$ layers.
Notice that the last layer consists of (some) leaves of $\cal T$ and
that $t=O(\log k)$, which was proved above ($k\le \lambda - 1$).

Let $Q$ be a node of $\cal T$ at layer $j$. We will prove that
\begin{equation}\label{ineq1}pw(UN(D[V(Q)]))< 2(t-j+2.5)k\end{equation} Since
$t=O(\log k)$, (\ref{ineq1}) for $j=1$ implies that  the underlying
undirected graph of $D$ is of pathwidth $O(k \log k)$.

We first prove (\ref{ineq1}) for $j=t$ when $Q$ is a path from the
decomposition. Let $W=(L(T)\cup S^+_{\ge 2}(T)\cup F(T))\cap V(Q),$
where $F(T)$ is the set of first vertices of maximal paths of $T$
consisting of link vertices. As in the proof of Theorem \ref{dagth},
it follows from Facts 1 and 2 that $|W|<4k.$ Obtain a digraph $R$ by
deleting from $D[V(Q)]$ all arcs  in which at least one end-vertex
is in $W$ and which are not arcs of $Q$. As in the proof of Theorem
\ref{dagth}, it follows from Lemma \ref{char1ae} and 1-AE optimality
of $T$ that there are no forward arcs for $Q$ in $R$. Let $Q=v_1v_2
\dots v_q$. For every $j\in [q]$, let
 $V_j = \{v_i:\ i\in [j]\}$. If for some $j$ the set $V_j$
contained $k$ vertices, say $\{v_1',v_2',\cdots ,v_k'\}$, having
in-neighbors in the set $\{v_{j+1},v_{j+2}, \dots, v_q \}$, then $D$
would contain an out-tree with $k$ leaves formed by the path
$v_{j+1}v_{j+2} \dots v_q$ together with a backward arc terminating
at $v_i'$ from a vertex on the path for each $1\leq i \leq k$, a
contradiction. Thus $vs(UN(D_2[P]))\leq k.$ By
Proposition~\ref{sovp_pw_vs}, the pathwidth of $UN(R)$ is at most
$k$.  Let $(X_1, X_2, \ldots, X_s)$ be a path decomposition of
$UN(R)$ of width at most $k$. Then $(X_1\cup W, X_2\cup W, \ldots,
X_s\cup W)$ is a path decomposition of $UN(D[V(Q)])$ of width less
than $k+4k.$ Thus,
\begin{equation}\label{ineq2}pw(UN(D[V(Q)]))<5k\end{equation}

Now assume that we have proved (\ref{ineq1}) for $j=i$ and show it
for $j=i-1$. Let $Q$ be a node of layer $i-1$. If $Q$ is a leaf of
$\cal T$, we are done by (\ref{ineq2}). So, we may assume that $Q$
has sons $Q'$ and $Q''$ which are nodes of layer $i.$ In the
$\beta$-decomposition of $T$ given before this theorem, we saw that
the vertices of $T'$ have at most $\lambda+1$ out-neighbors in $T''$
and the vertices of $T''$ have at most $\lambda+1$ out-neighbors in
$T'$. Similarly, we can see that (in the $\beta$-decomposition of
this proof) the vertices of $Q'$ have at most $k$ out-neighbors in
$Q''$ and the vertices of $Q''$ have at most $k$ out-neighbors in
$Q'$ (since $k\le \lambda - 1$). Let $Y$ denote the set of the
above-mentioned out-neighbors on $Q'$ and $Q''$; $|Y|\le 2k.$ Delete
from $D[V(Q')\cup V(Q'')]$ all arcs in which at least one end-vertex
is in $Y$ and which do not belong to $Q'\cup Q''$

Let $G$ denote the obtained digraph. Observe that $G$ is
disconnected and $G[V(Q')]$ and $G[V(Q'')]$ are components of $G$.
Thus, $pw(UN(G))\le b$, where
\begin{equation}\label{ineq3}b=\max\{pw(UN(G[V(Q')])),pw(UN(G[V(Q'')]))\}<
2(t-i+4.5)k\end{equation} Let $(Z_1,Z_2,\ldots ,Z_r)$ be a path
decomposition of $G$ of width at most $b.$ Then $(Z_1\cup Y,Z_2\cup
Y,\ldots ,Z_r\cup Y)$ is a path decomposition of $UN(D[V(Q')\cup
V(Q'')]$) of width at most $b+2k<2(t-i+2.5)k.$ \qed
 \end{proof}
Similar to the proof of Corollary~\ref{cor:acyclic}, we obtain the
following:
\begin{corollary}\label{cor:mainresult}
For a strongly connected digraph $D$, the problem $k$-DMLOB can be
solved in time $2^{O(k\log ^2k)}\cdot n^{O(1)}$.
\end{corollary}

\section{Discussion and Open Problems}

In this paper, we  continued algorithmic and combinatorial
investigation of the {\sc Directed Maximum Leaf Out-Branching}
problem. In particular, we showed that for every strongly connected
digraph $D$ of order $n$ and with minimum in-degree at least 3,
$\ell_s(D)=\Omega(n^{1/3})$. The most interesting open combinatorial
question here  is whether this bound is tight. It would be even more
interesting to find the maximum number $r$ such that
$\ell_s(D)=\Omega(n^r)$ for every strongly connected digraph $D$ of
order $n$ and with minimum in-degree at least 3. It follows from our
results that $\frac{1}{3}\le r\le \frac{1}{2}.$

We also provided an algorithm of time complexity $2^{O(k\log^2
k)}\cdot n^{O(1)}$ which solves $k$-DMLOB for a strongly connected
digraph $D$. The algorithm is based on a combinatorial bound on the
pathwidth of the underlying undirected graph of $D$. Unfortunately,
this technique does not work on all digraphs. It remains an
algorithmic challenge to establish the parameterized complexity of
$k$-DMLOB on all digraphs.

Notice that $\ell(D)\ge \ell_s(D)$ for each digraph $D$. Let $\cal
L$ be the family of digraphs $D$ for which either $\ell_s(D)=0$ or
$\ell_s(D)=\ell(D)$. The following assertion shows that $\cal L$
includes a large number digraphs including all strongly connected
digraphs and acyclic digraphs (and, also, well-studied classes of
semicomplete multipartite digraphs and quasi-transitive digraphs,
see \cite{bang2000} for the definitions).

\begin{proposition}[\cite{alonLNCS4596}]\label{L} Suppose that a digraph $D$ satisfies
the following property: for every pair $R$ and $Q$ of distinct
strong components of $D$, if there is an arc from $R$ to $Q$ then
each vertex of $Q$ has an in-neighbor in $R$. Then $D\in \cal L$.
\end{proposition}

Let $\cal B$ be the family of digraphs that contain out-branchings.
The results of this paper proved for strongly connected digraphs can
be extended to the class ${\cal L}\cap {\cal B}$ of digraphs since
in the proofs we use only the following property of strongly
connected digraphs $D$: $\ell_s(D)=\ell(D)>0$.

For a digraph $D$ and a vertex $v$, let $D_v$ denote the subdigraph
of $D$ induced by all vertices reachable from $v.$ Using the
$2^{O(k\log^2 k)}\cdot n^{O(1)}$ algorithm for $k$-DMLOB on digraphs
in ${\cal L}\cap {\cal B}$ and the facts that (i) $D_v\in {\cal
L}\cap {\cal B}$ for each digraph $D$ and vertex $v$ and (ii)
$\ell(D)=\max\{\ell_s(D_v)| v\in V(D)\}$ (for details, see
\cite{alonLNCS4596}), we can obtain an $2^{O(k\log^2 k)}\cdot
n^{O(1)}$ algorithm for $k$-DMLOT on {\em all} digraphs. For acyclic
digraphs, the running time can be reduced to $2^{O(k\log k)}\cdot
n^{O(1)}$.




\end{document}